\documentclass[traditabstract]{aa}
\usepackage{epsfig}
\usepackage{graphicx}
\usepackage{epstopdf}
\usepackage{natbib}
\usepackage{amsfonts}
\usepackage{amsmath}
\bibpunct{(}{)}{;}{a}{}{,}

\begin{document}

% - new commands
\newcommand{\hi}{\ion{H}{i}~}
\newcommand{\hii}{\ion{H}{ii}~}

   \title{MOND and the dark baryons}

   \author{O. Tiret         \inst{1,2}       
\and          F. Combes          \inst{1}  
 }

%  \offprints{O. Tiret }

   \institute{  Observatoire de Paris, LERMA, 
            61 Av. de l'Observatoire, F-75014, Paris, France 
            \and
                     SISSA, via Beirut 4, I-34014 Trieste, Italy 
                       }

   \date{Received XXX 2008/ Accepted YYY 2008}

   \titlerunning{}

   \authorrunning{Tiret \& Combes}

\abstract{
We consider for the first time the implications on the modified gravity MOND model of galaxies,
of the presence of dark baryons, under the form of cold molecular gas in galaxy discs.
 We show that MOND models of rotation curves are still valid and universal, but the critical acceleration
$a_0$ separating the Newtonian and MONDian regimes has a lower value.  We quantify this modification,
as a function of the scale factor $c$ between the total gas of the galaxy and the measured atomic gas.
 The main analysis concerns 43 resolved rotation curves and allows us to find the best pair
($a_0 = 0.96\times 10^{-10}$ m.s$^{-2}$, $c = 3$), which is also compatible to the one obtained from a second method by
minimizing the scatter in the baryonic Tully-Fisher relation. }

%{$ $}{$ $}{$ $}{$ $}


\maketitle

%---------------------------------------------------------------

\section{Introduction} 

The missing mass problem at galactic scales is revealed by rotation curves of spiral galaxies,
 where stars and gas rotate with speeds higher than expected (e.g. Sofue \& Rubin, 2001), 
in the frame of Newtonian gravity. This missing mass has been called dark matter and 
has now been developed into a standard cosmological model $\Lambda$CDM (e.g. Blumenthal et al. 1984). 
Dark matter particles collapsed to form the first structures of the Universe, and it is in those potential wells 
that the baryonic gas infalls and cools to form galaxies. As a result, galaxies are embedded in a  
 spheroidal dark matter halo with a mass profile obeying $M(r)\propto r$ in the outer parts, so that
the Keplerian circular velocity, $v_c=\sqrt{GM(r)/r}$, tends to be constant in the majority of observed 
rotation curves  (e.g. Bosma 1981).  In the central parts however, the CDM numerical simulations
predict a cuspy dark matter profile, with a relative CDM amount much larger than what is observed
(Navarro et al 1997, Navarro \& Steinmetz  2000).

An alternative, without invoking any dark matter, is to consider a modification of the Newtonian gravity law 
at low accelerations, so that the luminous mass is sufficient to describe the dynamics of the 
gravitational system. This was done empirically by Moti Milgrom who proposed the MOdified Newtonian 
Dynamics (MOND) paradigm (Milgrom, 1983). The MOND regime is distinguished from the Newtonian 
regime when the characteristic acceleration of the gravitational system falls below a critical 
acceleration $a_0\sim 1.2\times 10^{-10}$ m.s$^{-2}$. In the deep MOND regime, the modification $a_M$ of the 
Newtonian acceleration, $a_N$, can be written as $a_M=\sqrt{a_0 a_N}$. The asymptotic circular 
velocity is thus a constant : $v^4=GMa_0$. This is the Tully-Fisher law, which finds there a
justification.

In this paper, we consider for the first time how the problem of missing baryons 
could be made compatible with the MOND phenomenology. In the cosmic baryon budget, it 
is now well known that only about 6\% of them shine as stars and gas in galaxies (Fukugita et al. 1998). 
Locally, UV absorption in front of background sources have shown that as much as 30\% of baryons
could be associated with the Lyman$\alpha$ forest 
and an uncertain fraction (5-10\%) with the warm-hot medium (WHIM at 10$^5$-10$^6$K,
Nicastro et al 2005, Danforth et al 2006). So about half of the baryons are not yet 
accounted for. Most of them should be in the cosmic filaments, in the intergalactic medium,
but it appears unavoidable that a significant fraction of them exist as cold gas in galaxies
themselves, as suggested by Pfenniger et al (1994), and Pfenniger \& Combes (1994). 
From the rotation curves, it is possible to put an upper limit on the
fraction of dark baryons in galaxies: the dark baryons cannot overpredict the rotation
speeds. HI interferometric studies of nearby galaxies have 
shown that the dark matter distribution necessary to fit the rotation curves 
of galaxies follows the surface density of the atomic gas (Bosma 1981). Hoekstra et al. (2001)
estimate that a factor of 7-10 between the dark baryons and atomic gas reproduces the rotation curve 
of any galaxy, giant or dwarf, high or low surface brightness, without 
adding any dark matter halo. This scale factor means than there cannot be more
than 10\% of the total baryons in the form of cold gas in galaxies.
The observed HI extended rotation curves roughly require only to double
the fraction of baryons in galaxies at maximum.  Out of the 50\% of the dark baryons
still not accounted for, no more than 10\% can be in galaxies, but they can be less abundant.

Another way to trace the dark baryons in galaxy discs is the Tully-Fisher relation. 
This observational law involves the luminosity versus rotational velocity, which is equivalent to the 
stellar mass versus the velocity (with a constant mass-to-light ratio). A divergence from this law was found 
for late-type galaxies. On the Tully-Fisher diagram, the stellar mass of these galaxies is too small
for the velocity observed. The problem was solved by taking into account the mass of the atomic gas 
instead of only the stellar mass (McGaugh et al., 2000). The next step was to suppose that all the gas
is not seen (dark baryons). By this approach, Pfenniger \& Revaz (2005) find that a factor of 3  between the total and atomic gas
reduces the scatter of the baryonic Tully-Fisher relation. More recently, Begum et al. (2008) carried out 
the same analysis but focusing on a sample with low-mass galaxies, which are more constraining because 
they generally contain more gas. They find a scale-factor of around 9.

From a dynamical point a view, Revaz et al. (2008) performed some numerical simulations of galaxies in Newton gravity with dark matter. They add 
a cold dark baryon component and show that a factor of up to 5 between the total and atomic gas is realistic 
to reproduce the global behavior of the galaxy: the stability against local axisymmetric collapse
is ensured through a low dissipation, and patterns still spiral can develop due to the
self-gravity of the disc.

It is thus interesting to take into account the possible existence of dark baryons 
in the framework of MOND at galactic scales. If some fraction of the assumed dark matter
is real, the actual critical acceleration $a_0$ is overestimated. We want to derive
the new possible limiting acceleration, and interpolating function $\mu$.
The contribution of dark baryons has been discussed in modeling galaxy cluster cores by Milgrom (2007). 
At galactic scales, 
Gentile et al. (2007) have studied the contribution of neutrinos to rotation curves.

The paper is organized as follows:  in section 2, we introduce the contribution of dark baryons at galactic scale, in section 3 we describe the galaxy sample. 
The modeling of galaxies is discussed in section 4. In section 5, the implications of the
existence of dark baryons 
with MOND are explored through rotation curve fits and the Tully-Fisher law, then discussed in section 6.

\section{Overview}
\label{db}

We consider that part of the missing mass problem in galaxies could be alleviated by the presence of some dark baryons. They could be in the form of cold molecular gas H$_2$ in the outer flaring disc. This scenario has been studied in the framework of Newtonian gravity (Pfenniger \& Combes, 1994; Combes \& Pfenniger, 1997). In this model the amount of non-baryonic DM condensed in galaxies is reduced, but still necessary.

\begin{figure}
 \centering
  \resizebox{\hsize}{!}{\includegraphics{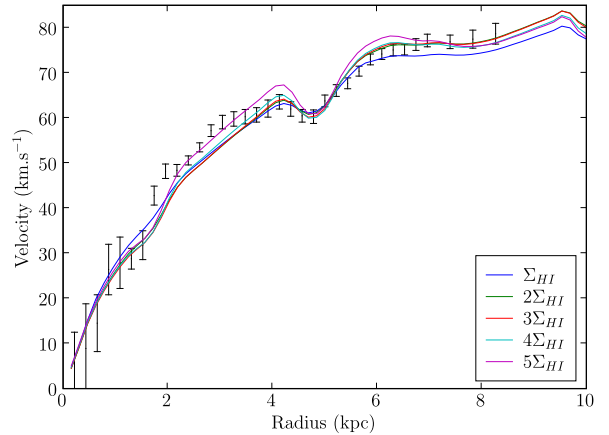}}
 \caption{Rotation curve fit of NGC 1560, in the MOND framework with dark baryons in the form
of cold molecular gas. If this molecular gas is such that mass $M_{gas}=cM_{at}$, the value of the actual estimation of the critical acceleration $a_0=1.2\times 10^{-10}$ m.s$^{-2}$ must be reduced (divided by a factor $c$, to a first approximation), to be in agreement with the observations. The factor $c$ is indicated in the bottom right corner, for each curve fitted.}
 \label{fig:CR_H2}
\end{figure}

At present, in the framework of modified gravity, since the dark molecular gas increases the baryonic mass, the critical acceleration of MOND will be
reached at a lower level, and this is not necessarily in agreement with a single acceleration $a_0$
for all galaxies. Is the presence of some dark baryons in galaxies still compatible with MOND? 
As the required "phantom" dark matter is proportional
to the HI gas (Hoekstra et al 2001), it appears quite possible to combine the presence of
dark baryons with MOND, by only reducing the value of the critical acceleration.
For example, we examine the dwarf LSB galaxy NGC 1560, dominated by dark matter in the Newtonian frame. Its rotation curve reaches a plateau at $~7$ kpc. In accordance with MOND, this maximum velocity can be written:
$$v_c^4=GMa_0.$$
The mass M, to a first approximation, corresponds to the mass of the atomic gas (HI+He) $M_{at}$, since
 at this outer radius, the stellar content is negligible.
If we consider that this galaxy may contain as much molecular gas as atomic gas, which means that the total mass is 
$M_{gas}=2 M_{at}$, the actual value $a_0$ must be divided by 2 to conserve the same asymptotic 
velocity. With this model, if the total mass of gas is equal to the atomic gas mass times a scale factor, 
$$M_{gas}=c M_{at},$$ the critical acceleration $a_0$ must be divided by $c$ to keep the same  asymptotic velocity.
It works correctely for the galaxy NGC1560, as shown in the Fig. \ref{fig:CR_H2}. The most remarkable is the shape of the rotation curve which is conserved by modifying the values $(c,a_0)$, not just the maximum velocity. The wiggle at 5 kpc is well reproduced until $c=5$. Beyond this value, the oscillation is more and more pronounced and  no longer matches the data.

It might be surprising that two very different models succeed in representing equally well
the rotation curves of galaxies: the first is to assume the presence of 7-10 times more gas than 
observed in HI, the second is to modify the gravity. For dwarf irregular galaxies, where the stellar
component is minor, the visible matter is dominated by the  HI gas
and the total mass is dominated by the dark matter. 
With these approximations, we can write two equations: in the Newton model,
the acceleration at the outskirt of the disk is  $V_{rot}^2/r \sim 7 G M(HI)/r^2$,
since the total mass is very close to 7 times the gas mass for these dwarfs, and in the MOND model,
$V_{rot}^2/r \sim ( a_0 G  M(HI)/r^2)^{1/2}$, since they are in the deep MOND regime in the outer
parts. Combining these two relations implies that the acceleration in these dwarfs is always 
about 7 times less than $a_0$, which is remarkable (cf also Fig \ref{fig:mass_acc_discr}).
As for the radial distribution, even for the dwarf galaxies, the regime is still partially Newtonian 
in the center, and the two models work equally well, because of a combination of 
factors including the stellar component and the interpolation MOND function
(see Figure 1).

In this paper, we study whether MOND is still consistent with the presence of dark baryons such as the molecular gas H$_2$. Is there still a critical acceleration universal for all the galaxies? What is the proportion of dark baryons that is consistent with the observations, in the framework of MOND? We try to answer these questions in two ways. First we analyze a \textit{first sample} of galaxies by fitting their rotation curves; this is the same approach as Hoekstra et al. (2001). The second method consists of
optimizing the scatter in the baryonic Tully-Fisher relation, as in Pfenniger \& Revaz (2005) and Begum et al. (2008), but interpreted in a modified gravity framework (see sect. \ref{sec:tfrelation}). We do this using the \textit{first} and the \textit{second sample}.

\section{Galaxy sample}

\begin{figure}
 \centering
 \resizebox{\hsize}{!}{
 \includegraphics{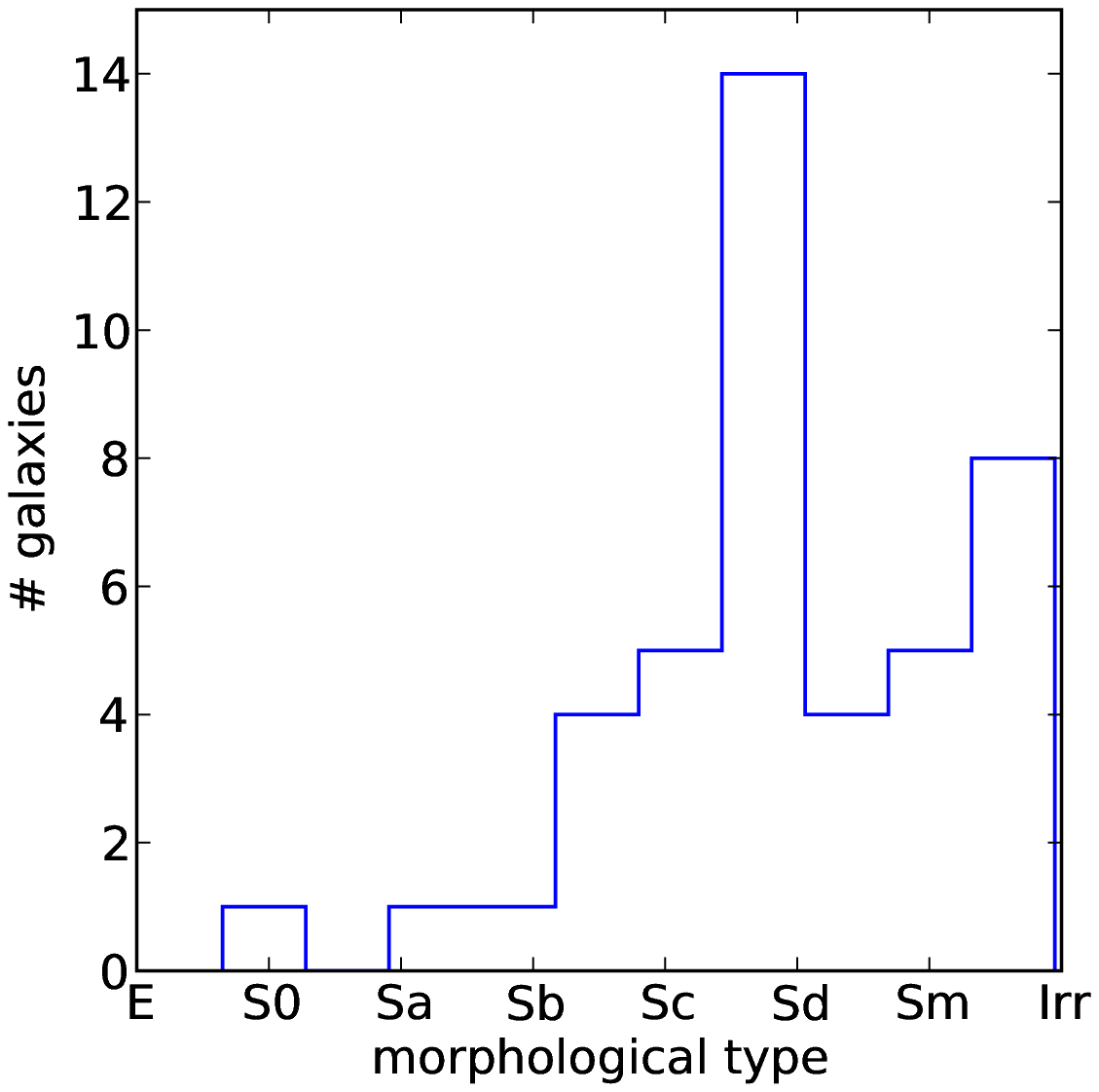}
\includegraphics{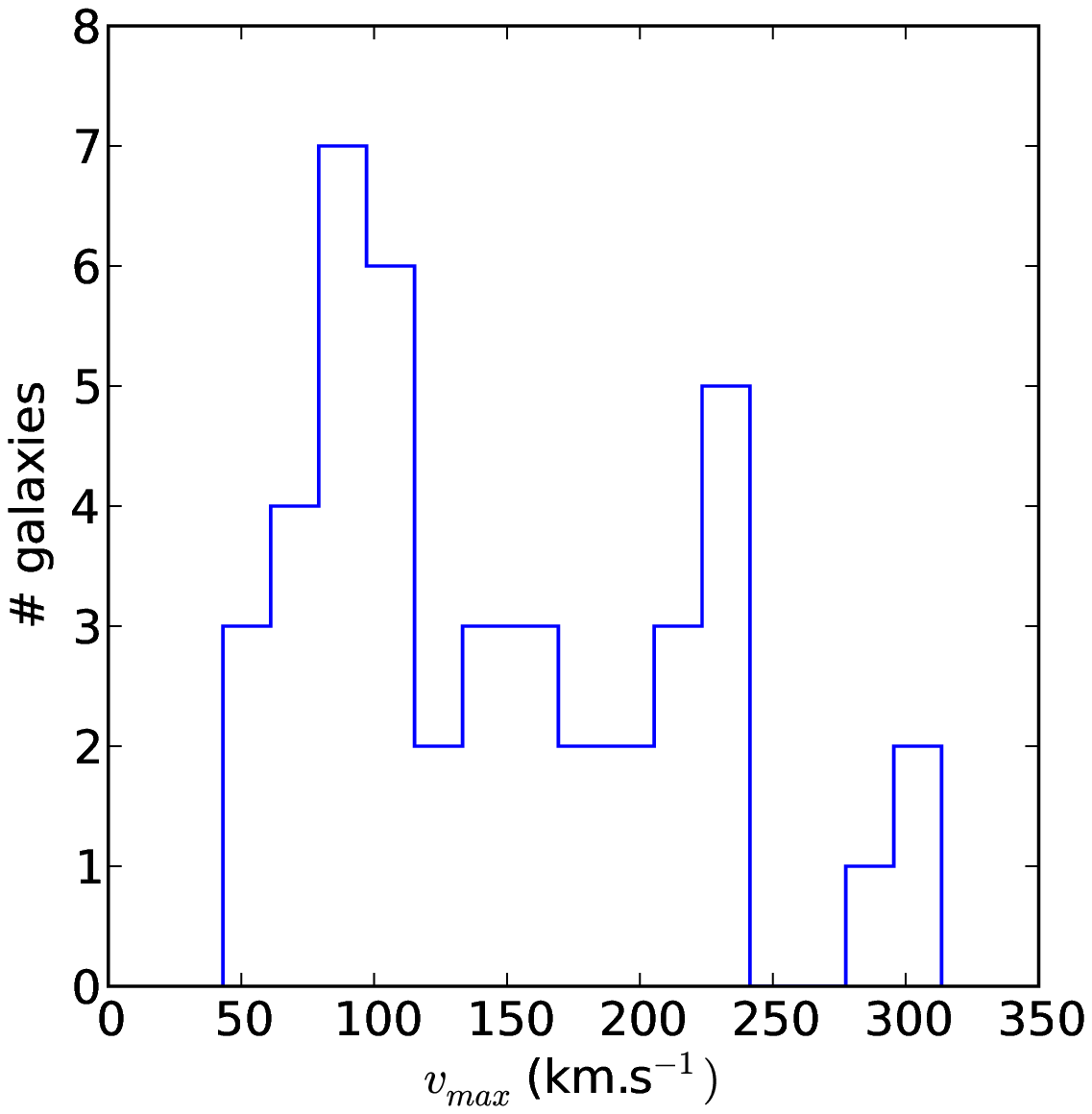}}	
 % ty.eps: 1048592x1048592 pixel, 300dpi, 8878.08x8878.08 cm, bb=
 \caption{\textit{Left}: Morphological type in the sample made of 43 nearby galaxies. \textit{Right}: Distribution of the maximum velocity in the galaxy rotation curves.}
\label{fig:sample}
\end{figure}

\subsection{First sample}

We compiled from the literature a galaxy sample of 43 nearby objects (Table \ref{tab:samp}).
For each of them, the resolved HI kinematical data are available. They are unperturbed 
(without large asymmetries such as lopsidedness or tidal interactions) so that a reliable rotation curve can be obtained. The galaxies have inclinations on the plane of the sky that vary from $30^{\circ}$ to $90^{\circ}$. A stellar light profile is also
available in the literature, so that the contribution of the visible mass can
be computed with reasonable accuracy.

We have searched for late-type galaxies, where the fraction of gas (and dark matter) 
is more important. In particular we gathered a large number of dwarf irregulars,
which are dominated by dark matter. Therefore, they are a convenient laboratory to determine
the radial dark matter distribution, independently of the stellar mass-to-light ratio. 
Fig.\ref{fig:sample} (left) displays a histogram of the morphological type in the sample; 70$\%$ of the galaxies are late-type. Fig. \ref{fig:sample} (right) shows the distribution of the maximum velocity of galaxies; the median of the sample is $V_{max}=128$ km.s$^{-1}$.

\begin{table*}
\centering
\begin{tabular}{lrcclrcccl}
\hline
\hline
Name & D        & $L_B$             & $M_{HI}$             &Type& $V_{max}$      & Inclination &  $r_{\star}$ &  $rmax/r_{\star}$  &References  \\
           & (Mpc)  &  ($L_{\odot}$)&  ($M_{\odot}$)     &          &   (km.s$^{-1}$)&     (degree)&    (kpc)         &                                &              \\
\hline

DDO154	&   4.0	&   3.4E07  &  1.8E08      			&I&  43 	& 32 &   0.5   &    15   & (1)            \\
DDO168	&   4.0	&   2.2E08  &  1.8E08       			&I&  52 	& 80 &   0.5   &      8   & (2)(3)         \\
DDO170	&  15.8	&   3.1E07  &  1.7E09       			&I&  66 	& 80 &   0.6   &    19   & (4)            \\
IC2574	&   2.5 	&   5.4E08  &  2.1E08    			&SABm&  67 	& 80 &   1.1   &      6   & (5)  (31)          \\
M33	&   0.8 	&   2.8E09  &  1.8E09      			&Sc& 134 	& 55 &   1.7   &      9   & (6)(7)        \\
NGC1003	&  11.8 	&   5.7E09  &  2.5E09     			&Sc& 112 	& 81 &   2.6   &    11   & (2)(31)            \\
NGC1560	&   3.0 	&   6.6E08  &  7.0E08     		&Scd&  78 	& 83 &   0.9   &      8   & (8)            \\
NGC2366	&   2.6 	&   3.4E08  &  3.4E08    		&IB&  45 	& 90 &   0.8   &      7   & (9)            \\
NGC2403	&   3.6 	&   5.9E09  &  3.6E09    			&SABc& 140& 60 &   2.9   &      6   & (10)(11)(12)   \\
NGC24	&   6.8 	&   1.1E09  &  3.8E08     			&Sc& 109     & 70 &   1.1   &      9   & (13)           \\
NGC247	&   2.8 	&   2.2E09  &  8.0E08    			&SABc& 107     & 75 &   1.9   &      5   & (14) (15)      \\
NGC253	&   2.6 	&   1.2E10  &  6.6E08    			&SABc& 224 	& 78 &   2.1   &      4   & (16)         \\
NGC2841	&  20.0 	&   4.9E10  &  2.1E10     			&Sb& 292 	& 68 &   3.3   &    26   & (10)(11)       \\
NGC2903	&  10.4 	&   2.5E10  &  5.8E09    			&SABb& 181 	& 56 &   2.9   &    13   & (10)(11)       \\
NGC2915	&   5.6 	&   4.7E08  &  7.7E08     			&SBa&  93 	& 63 &   0.4   &    42   & (17)           \\
NGC2998	&  70.7 	&   3.9E10  &  2.8E10    			&SABc& 200 	& 61 &   5.0   &    10   & (2) (18)       \\
NGC300	&   2.9 	&   3.5E09  &  1.4E09     			&Scd&  93 	& 39 &   2.6   &      6   & (19)           \\
NGC3109	&   1.9 	&   9.8E08  &  3.8E08     			&SBm&  67 	& 80 &   1.0   &      9   & (11)          \\
NGC3198	&  11.6 	&   1.1E10  &  7.3E09     		&Sc& 149 	& 70 &   2.9   &    12   & (11)(20)      \\
NGC3726	&  14.9 	&   1.3E10  &  7.6E09      	 		&Sc& 160 	& 49 &   3.7   &      8   & (9)           \\
NGC4203	&  15.1 	&   3.9E09  &  7.5E08   			&E-SO& 152 	& 90 &   2.3   &      7   & (9)           \\
NGC4242	&   7.3 	&   1.3E09  &  4.5E08      			&Sd&  98 	& 51 &   1.5   &      7   & (9)           \\
NGC4258	&   7.3 	&   1.7E10  &  3.7E09    			&SABb& 216 	& 72 &   2.5   &    11   & (9)  \\
NGC4395	&   2.7 	&   6.4E08  &  4.7E08      		&Sm&  85 	& 90 &   1.5   &      4   & (9)  \\
NGC45	&   5.9 	&   1.2E09  &  1.3E09    			&SABd& 100 	& 60 &   2.0   &      8   & (13) \\
NGC4725	&  11.9 	&   1.5E10  &  2.6E09    			&SABa& 223 	& 54 &   4.3   &     6    & (9)  \\
NGC5033	&  18.4 	&   2.5E10  &  1.5E10     			&Sc& 196 	& 66 &   5.5   &     8    & (10)(18)  \\
NGC5055	&  10.3 	&   2.3E10  &  9.8E09     			&Sbc& 172 	& 56 &   3.8   &   12    & (9)\\
NGC5371	&  40.0 	&   5.6E10  &  9.9E09     			&Sbc& 313 	& 54 &   7.4   &     5    & (9) (10) \\
NGC55	&   1.6 	&   4.6E09  &  6.8E08     			&SBm&  86 	& 85 &   1.4   &     7    & (21)  \\
NGC5533	&  58.3 	&   3.5E10  &  3.0E10     			&Sab& 230 	& 60 &   5.5   &   14    & (2)  (22)  (23)\\
NGC5585	&   7.6 	&   1.9E09  &  1.2E09    		&SABc&  89 	& 53 &   1.7   &     6    & (24)  \\
NGC5907	&  11.0 	&   1.3E10  &  2.4E09      			&Sc& 219 	& 87 &   2.4   &   13    & (25)  \\
NGC6503	&   4.8 	&   2.0E09  &  1.0E09     	 		&Sc& 122 	& 74 &   1.1   &   15    & (2) (9) \\
NGC6674	&  51.8 	&   3.4E10  &  2.7E10      			&Sb& 240 	& 62 &   5.5   &   13    & (2)  \\
NGC6946	&   6.7 	&   1.8E10  &  7.8E09    			&SABc& 159 	& 31 &   3.4   &     5    & (26)  \\
NGC7331	&  12.8 	&   2.8E10  &  7.3E09     			&Sbc& 241 	& 75 &   2.2   &   14    & (2)(11)  \\
NGC7793	&   4.1 	&   2.8E09  &  1.0E09     			&Scd&  90 	& 53 &   1.5   &     5    & (27)  \\
NGC801	&  84.0	&   5.1E10  &  1.9E10      			&Sc& 218 	& 85 &   2.9   &   20    & (2) (18) \\
NGC925	&   6.5 	&   4.1E09  &  2.5E09     			&Scd& 114 	& 61 &   2.2   &     7    & (9) \\
UGC128	&  56.4 	&   1.0E09  &  6.2E09      			&Sd& 128 	& 32 &   5.0   &     7    & (28)  \\
UGC2259	&  10.0 	&   2.3E08  &  3.7E08     		&Sbd&  90 	& 53 &   1.1   &     7    & (11) (31) \\
UGC2885	&  84.0 	&   5.6E10  &  5.9E10     			&Sc& 298 	& 62 &11.5    &     6    & (18)  \\
\hline                                              
\end{tabular}                                       
                                                    
\caption{(1) Carignan \& Beaulieu (1989); (2) Broeils (1992a); (3) Sanders (1996); (4) Lake et al. (1990 ); (5) Martimbeau et al. (1994); (6) Newton (1980); (7) Corbelli (2003); (8) Broeils (1992b); (9) Wevers (1984); (10) Begeman (1987); (11) Kent (1987); (12) Sicking (1997); (13) Chemin et al. 2006; (14) Carignan \& Puche (1990b); (15) Carignan (1985); (16) Puche et al. (1991); (17) Meurer et al. (1994); (18) Kent (1986); (19) Puche et al. (1990); (20) Begeman (1989); (21) Puche et al. (1991a); (22) Broeils \& Knapen (1991); (23) Kent (1984); (24) Cote et al.(1991); (25) Sancisi \& van Albada (1987); (26) Carignan et al. (1990); (27) Carignan \& Puche (1990a); (28) de Blok et al. (1985); (29) Carignan et al. (1988); (30) Roelfsema \& Allen (1985); (31) Leroy et al.  2005}
\label{tab:samp}    

\end{table*} 

\subsection{Second sample}

A second sample of galaxies was build from the GoldMine database. It contains 576 galaxies selected with $v_c<$ 100 km.s$^{-1}$.
These galaxies are low-mass galaxies with a high gas fraction (as in Begum et al,2008). Unlike the first sample, they
are not kinematically resolved thus the stellar mass cannot be determined from a rotation curve fit. However the GoldMine database provides the stellar luminosity
in the K-band and the color B-V from which a mass to light ratio can be derived (Bell \& de Jong, 2001). It also gives the 
width of the HI line obtained by averaging the value at $20\%$ of the peak flux with the one at $50\%$ of the mean flux and the total mass in HI (see the GoldMine website).
\section{Method}

\subsection{Rotation curve analysis}
\label{sec:rca}
We analyzed the 43 galaxies (\textit{first sample}) where a complete set of data is available: the luminosity profile, 
the HI profile and the HI kinematics. The atomic gas mass is corrected for the primordial helium abundance
 ($M_{He}=0.3M_{HI}$). The mass-to-light ratios are estimated from the rotation curve fits in MOND, 
by taking into account the stellar and atomic disc only (fixing $a_0=1.2\cdot 10^{-10}$ m.s$^{-2}$). Note that in MOND when the baryonic
profile is determined the mass-to-light ratio is the only parameter of the fit, since the critical acceleration $a_0$ and the interpolating function $\mu$ are the same
whatever the considered galaxy. These mass-to-light ratios are given in Table \ref{tab:param}, 
the stellar mass-to-light ratios, in the B-band, vary between 
1 and 6 from the late-type to early-type galaxies.

The method consists of modeling all the rotation curves to find which pair  $(a_0,c)$ best fit 
each galaxy by minimizing the $\chi^2$. Then we look at the histogram of $a_0$ and $c$ to see if there is a value common to 
all the sample.

If MOND is still correct we expect to obtain a common value of $a_0$ for all the galaxies. On the other hand, the scale-factor c 
has no reason to be universal. Its value depends on the history of each galaxy because of interactions, mergers,  and the star formation rate.
But the mean value should be consistent with previous works (Hoekstra et al. 2001, Pfenniger \& Revaz 2005, Begum et al. 2008)

\subsubsection{Galaxy modeling}

The stellar disc is modelled by an exponential surface density:
$$\Sigma_\star(r)=\Sigma_{0\star}\exp(-r/r_\star),$$
\noindent and the bulge is represented by an Hernquist profile, of characteristic scale $r_b$:
$$ \rho(r)=\dfrac{8\rho(r_b)}{(r/r_b)(r/r_b+1)^3}. $$
\noindent The parameters $\Sigma_{0\star}$, $r_\star$, and $r_b$ are deduced from the luminosity profile (B-band) fits.

The HI gas surface density is taken directly from observations without any particular modeling. In this way, it is particularly constraining to fit the wiggles of rotation curves associated with gas overdensity.

\subsubsection{Rotation curve modeling}
The rotation curves of each component are computed 
using the Bessel functions. For a given surface density $\Sigma$, the Newtonian circular velocity can be written (Binney \& Tremaine, 1994):
$$v_c^2(r) = -r \int_0^{\infty} S(k) J_1(kr) kdk,$$
\noindent with
$$S(k) = -2\pi G \int_0^{\infty} J_0(kr) \Sigma(r)rdr.$$
$J_1$ and $J_2$ are Bessel functions of order 0 and 1.
For an exponential disc, it can be simplified to:
$$v_c^2(r) = 4 \pi G \Sigma_{0\star} r_\star y^2 \left[ I_0(y)K_0(y)- I_1(y)K_1(y)  \right],$$
\noindent where $y=r/2r_\star$. $I$ and $K$ are the modified Bessel functions.

The MOND rotation curves are deduced from the Newtonian acceleration, using the MOND formula:
$$a_N=a_M \mu(a_M/a_0)$$
and inversely,
$$a_M=a_N \nu(a_N/a_0)$$
\noindent When not explicitly noted otherwise,
we use the standard $\mu$-function, $\mu(x)=x/\sqrt{1+x^2}$, with $x=a_M/a_0$.

\subsection{Tully-Fisher relation}

\label{sec:tfrelation}

We analyzed the baryonic Tully-Fisher relation using the \textit{first} and \textit{second sample}, which are independent of that of 
McGaugh et al. (2000) and Begum et al. (2008). The method consists of representing $v^n \propto M$ with $M=M_{\star} + cM_{at}$ and determines
which c optimizes the scatter of the Tully-Fisher relation. The only difference compared to the analysis of the previous authors is that we interpret this
Tully-Fisher relation in MOND so we fix the slope to $n=4$ to avoid too many parameters in the fit.

Contrary to the rotation curve analysis, this one is global. This means that we can derive a mean scalefactor $c$ between the total gas and atomic gas mass for all the galaxies and not for an individual galaxy. Thus the scatter will never be zero. Then when the best value of c for the sample is obtained, the critical acceleration is simply given by the vertical offset of the relation:  $\log(v^4)= \log(M) + \log(Ga_0)$.

For the \textit{first sample} the velocity v plotted for the Tully-Fisher relation is the last point of the rotation curve. The stellar and gas mass is the one computed for the rotation curve analysis, integrated until the last point of the rotation curve.

For the  \textit{second sample}, the velocity is obtained from half the width of the HI line given in the GoldMine database. This velocity is corrected for the 
inclination (i) of the galaxy estimated by the ratio of the short and long axis, $$v=0.5 W/\sin(i).$$
The stellar mass is computed from the luminosity in the K-band. The mass-to-light ratio is given by the color B-V (Bell \& de Jong, 2001).

\subsection{The visible molecular gas}
  In about a dozen galaxies of the sample, there is information on the visible H$_2$ gas,
traced by the CO emission. It is observed in the more massive galaxies or early-types,
due to their higher metallicity and consequently greater abundance of CO.
In these galaxies, where the visible molecular component is a significant fraction of the
gaseous mass,   the radial distribution of the CO emission is exponential, with about the same
scale length as the optical exponential disk (e.g. Young \& Scoville 1991).
This radial distribution is completely different to the HI distribution, which is more
extended, with a surface density varying nearly with the inverse of the radius.
 In our approach, we therefore include the visible H$_2$ gas in the stellar component;
in these massive and early-type galaxies, it is never greater than 10\% of the stellar mass,
and enters in the uncertainty on the stellar mass to light ratio.

\section{Results}

\subsection{The critical acceleration}
\label{sec:acc_scale}

\begin{figure}
 \centering
  \resizebox{\hsize}{!}{
 \includegraphics{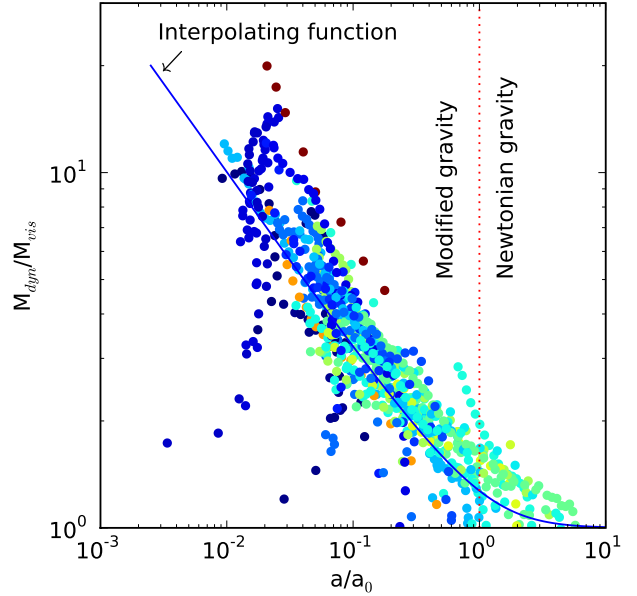}
}
 \caption{In Newtonian gravity, the dynamical to visible mass ratio is plotted versus  the acceleration of the visible matter. At large
acceleration, there is no need for dark matter. The MOND phenomenology is dictated by the rotation curves, which establish a strong correlation between the mass discrepancy and the acceleration. The gray (color) scale codes the type of the galaxies. Late-type galaxies are in dark gray (blue) while early-type are in light gray (yellow). The solid line (blue) represents the $\nu$ function associated with the standard $\mu$ function (see text).}
 \label{fig:mass_acc_discr}
\end{figure}

In this section, we discuss the dependence of the missing mass problem,
as a function of scale, on the acceleration.
Let us reason here in the context of Newtonian gravity. 
The process of rotation curve fitting determines the best mass-to-light ratio,
thus the visible mass profile
of the different components (stellar disc, bulge, gas):
$$M_{vis}=M_\star + M_b + M_g$$
\noindent and the dynamical mass is defined by 
$$M_{dyn}= M_{vis} + M_{DM}.$$

\noindent Let consider that the dark matter has a spherical symmetry:
$$v_c^2(r)-v_{vis}^2(r)=\dfrac{GM_{DM}(r)}{r}.$$

When the ratio $M_{dyn}/M_{vis}$ is plotted versus the acceleration (the Newtonian acceleration 
of the visible component), a very strong correlation is observed (Fig. \ref{fig:mass_acc_discr}). Whatever the galaxy considered, 
the amount of dark matter required to fit the rotation curve is the same for a given acceleration 
of the visible component. The color scale indicates the morphological type of the galaxy. Late-type galaxies are in dark gray (blue) while early-types tend towards light gray (yellow). This does not depend much on the estimation of the mass-to-light ratio $\Gamma_\star$, as shown by late-type galaxies which are gas dominated.This relation has also been found by McGaugh (2004), when estimating the mass-to-light ratio in several ways (maximum disc, stellar populations,
 ...). Whatever the method, a strong correlation between the mass discrepancy and the Newtonian acceleration is observed.

\noindent  This relation can be interpreted in two ways:
\begin{itemize}
 \item either we consider the context of Newtonian gravity and must find why 
baryons fall in dark matter haloes so that this relation with the acceleration of the visible matter 
is verified. The $\Lambda$CDM model does not give a real clue to this correlation. 
Van den Bosch \& Dalcanton (2000) try to obtain this relation using semi-analytical models of 
galaxy formation, in the $\Lambda$CDM framework. They must then consider star formation feedback effects 
through supernovae explosions, to reproduce the lack of high surface brightness dwarf galaxies.
They however do not succeed in eliminating massive galaxies with low surface brightness. 
They also tuned their galaxies to fit the observed Tully-Fisher relation in the $\Lambda$CDM model,
while galaxies automatically reproduce it in the MOND model.
It should be interesting to look at this in self-consistent cosmological simulations with baryons.
 \item we consider the alternative model of MOND, which is a direct application of the mass 
discrepancy-acceleration relation. In this frame, the visible mass is the only one,
and the gravitation law must be modified to remove the need for dark matter, such that 
$M_{dyn}$/$M_{vis}=1$. 
\end{itemize}

The mass discrepancy-acceleration relation is:

$$ {{M_{dyn}}\over{M_{vis}}} = f({{a_{vis}^{N}}\over{a_0}}). $$

\noindent The mass can be expressed in terms of acceleration:

$$ {{GM_{dyn}}\over{r^2}} = {{GM_{vis}}\over{r^2}}f\left({{a_{vis}^{N}}\over{a_0}}\right) $$
$$  a_{vis+DM}^{N} =a_{vis}^{N} f\left({{a_{vis}^{N}}\over{a_0}}\right).$$

A test particle should feel a MONDian acceleration,  $a^{M}$ equivalent to the Newtonian acceleration with dark matter, $a_{vis+DM}^{N}$ then,
$$  a^{M}=a_{vis}^{N} f\left({{a_{vis}^{N}}\over{a_0}}\right)$$

\noindent which is equivalent to Milgrom's formulation of MOND:
$$  a^{M}=a_{vis}^{N} \nu\left({{a_{vis}^{N}}\over{a_0}}\right)$$
\noindent or
$$  a_{vis}^{N}=a^{M} \mu\left({{a^{M}}\over{a_0}}\right),$$

\noindent with
$$ I(x)=x\mu(x) $$
$$\nu(y)=I^{-1}(y)/y.$$

Thus, rotation curves of galaxies tell us how Newtonian gravity should be modified to obtain a 
ratio $M_{dyn}/M_{vis}$ of one, without any dark matter halo. The $\mu$-function of MOND and the critical acceleration $a_0$ can be observationally derived from a Newtonian interpretation of the rotation curves.
The solid line in Fig. \ref{fig:mass_acc_discr}  represents the function: 
$$\nu(x)=   \sqrt{0.5 +0.5\sqrt{1+(2/x)^2}},$$ associated with the standard $\mu$-function. 
It is in good agreement with the data.

Let us note that for any model devised to fit rotation
curves (such as DM and MOND models), it is sufficient to verify the global Tully-Fisher relation,
together with the virial theorem, to reveal the mass discrepancy-acceleration relation.
The TF relation writes as $M_{vis} \propto V^4$, the virial equilibrium of galaxies
as $V^2 \propto M_{dyn}/R$; from the combination of these two relations, it
can be derived that $M_{dyn}/M_{vis} \propto R/V^2 \propto a_0/a$. Although these are
only global relations, they yield almost the correct slope, obtained for all galaxies,
including points internal to each galaxy.
The main difference here is that the TF relation is an integral part of the MOND model,
while it is fine tuned in the DM+Newton model.

\begin{figure*}[t]
 \centering
 \resizebox{\hsize}{!}{
 \includegraphics{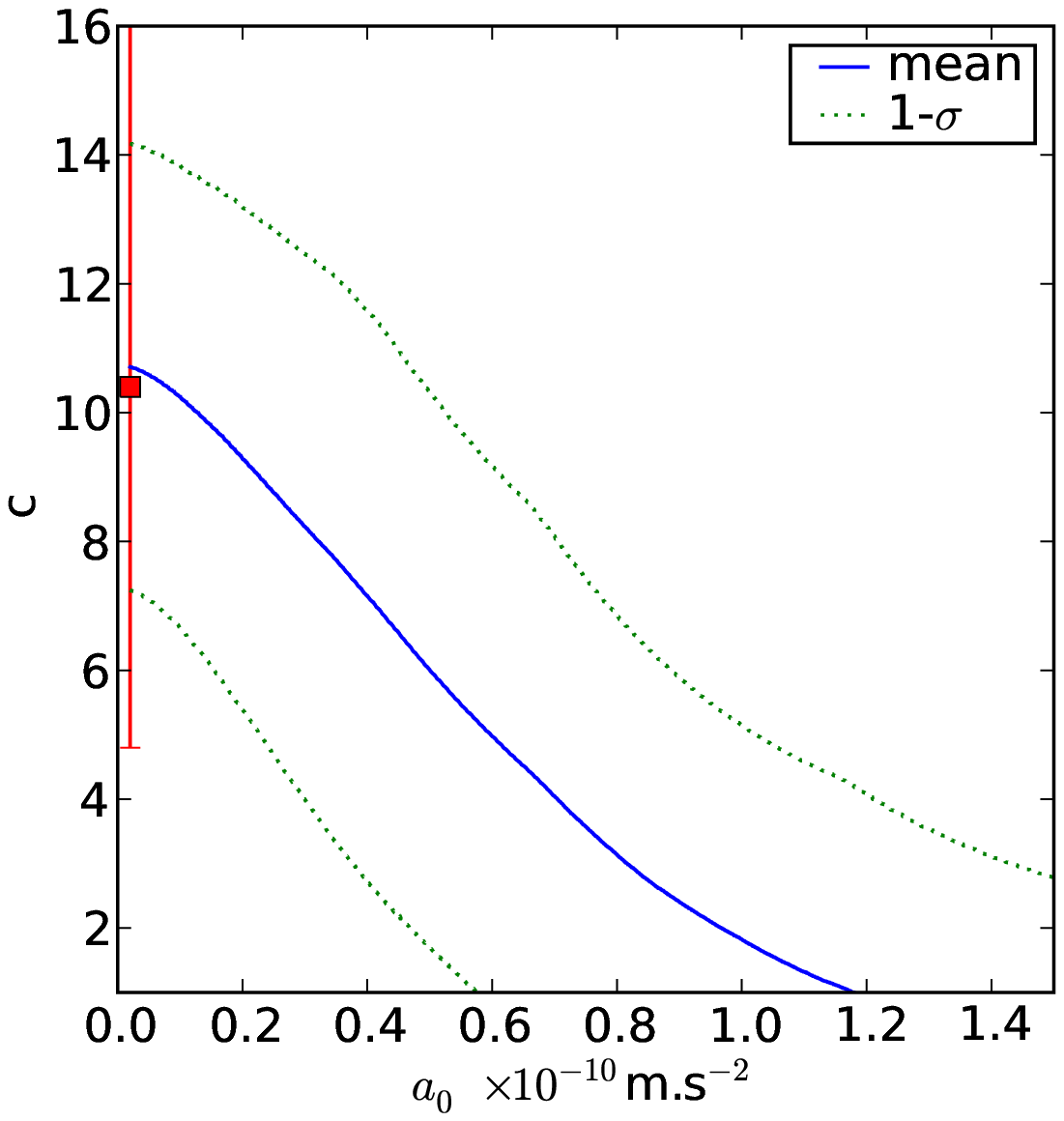}
 \includegraphics{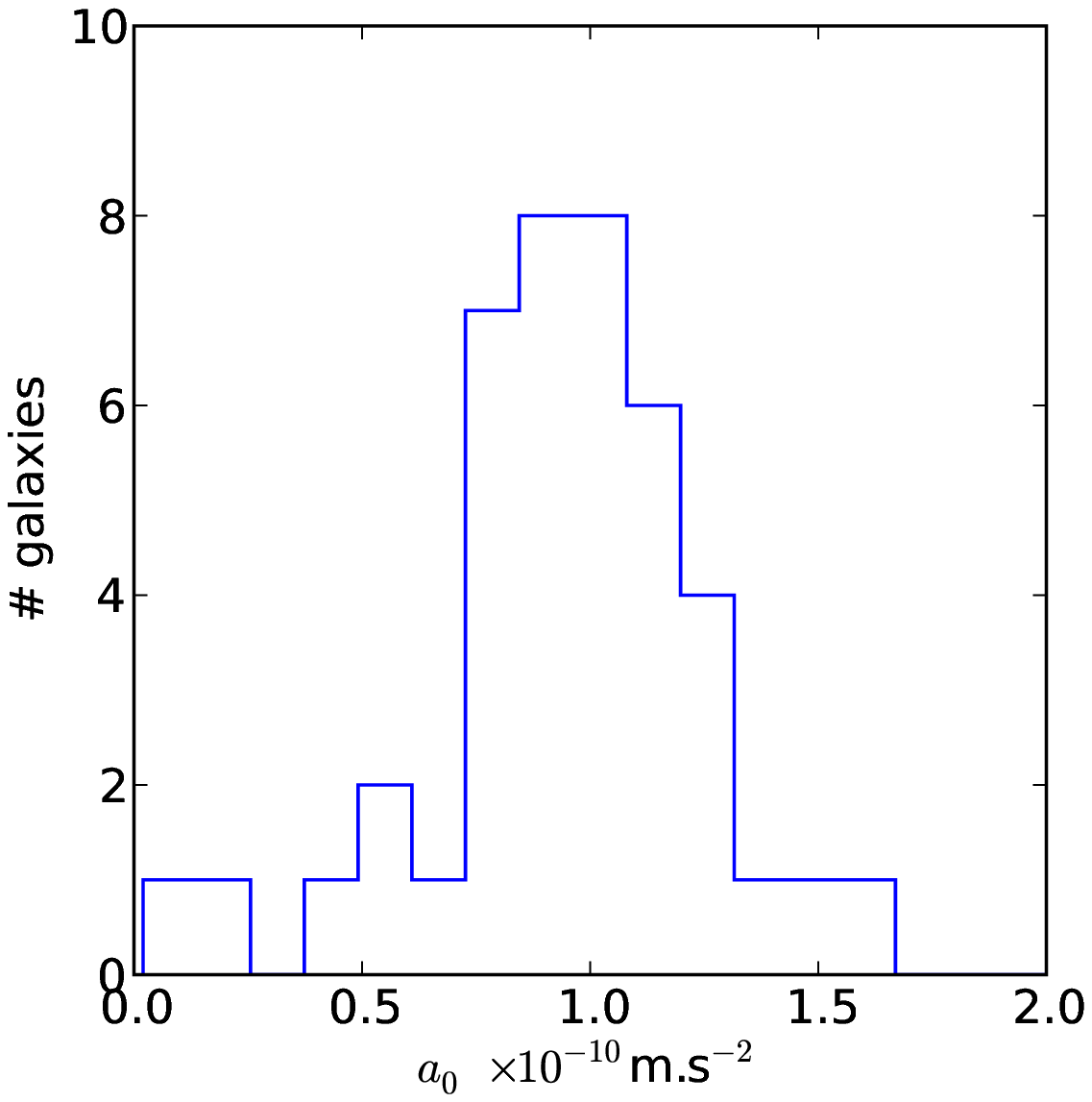}
\includegraphics{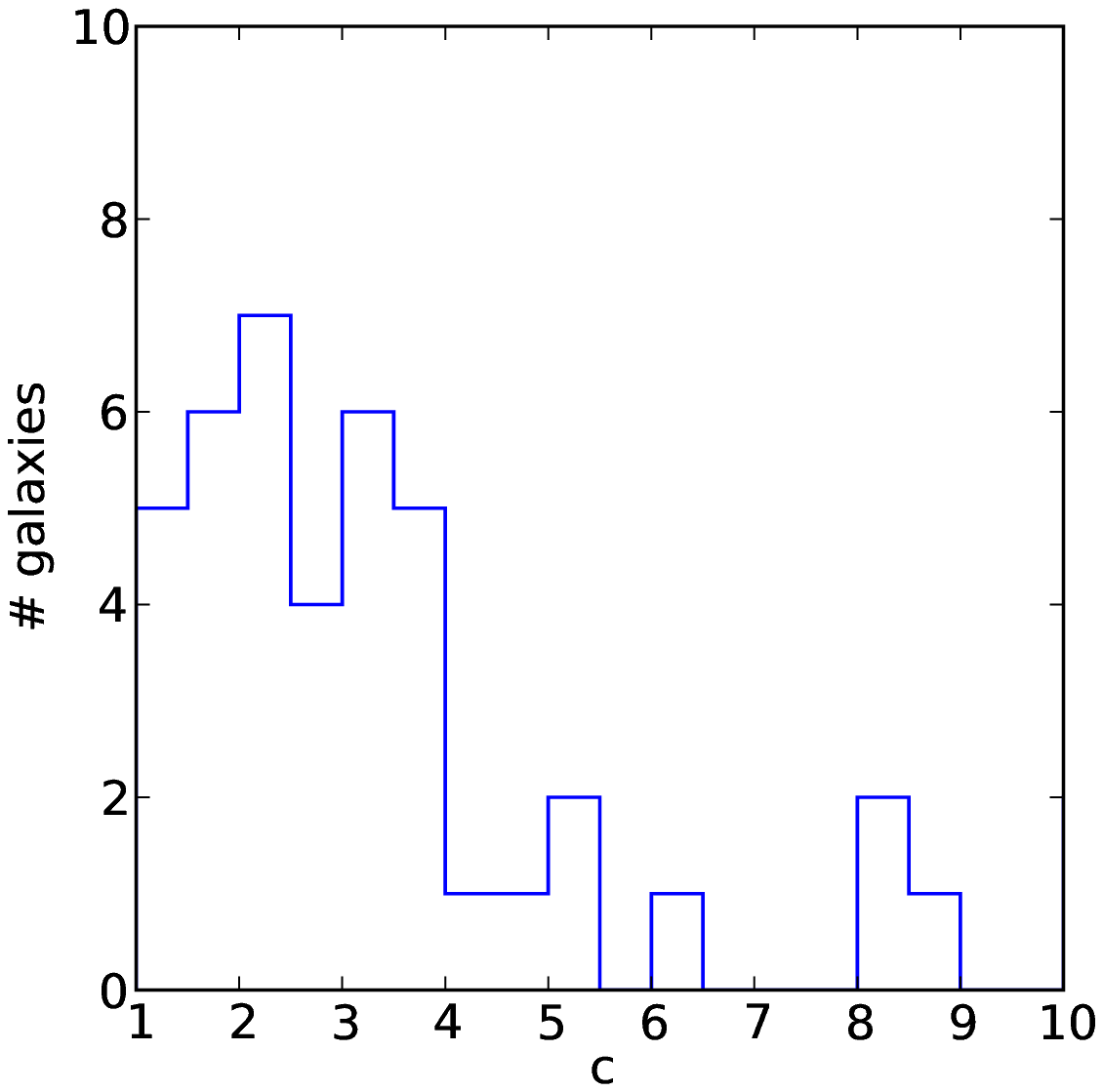}
}	
 % ty.eps: 1048592x1048592 pixel, 300dpi, 8878.08x8878.08 cm, bb=
 \caption{\textit{Left}: Variation of the mean c with $a_0$. The limit $a_0 \rightarrow 0$ corresponds to a Newtonian gravity without dark matter. The error bar (red) is the result from Hoekstra et al. (2001). When  $a_0 \rightarrow 1.2 \times 10^{-10}$ m.s$^{-2}$, the actual value of $a_0$, no additional baryons are required ($c\rightarrow1$). \textit{Middle}: Histogram of $a_0$ obtained when fitting the rotation curve sample including a dark baryon component. The universal critical acceleration peaks at  $a_0=0.96\times 10^{-10}$ m.s$^{-2}$. \textit{Right}: Histogram of c when all the rotation curves are fitted fixing  $a_0=0.96\times 10^{-10}$ m.s$^{-2}$. }
\label{fig:rc_ca0}
\end{figure*}

\begin{figure*}[t]
 \centering
 \resizebox{\hsize}{!}{
 \includegraphics{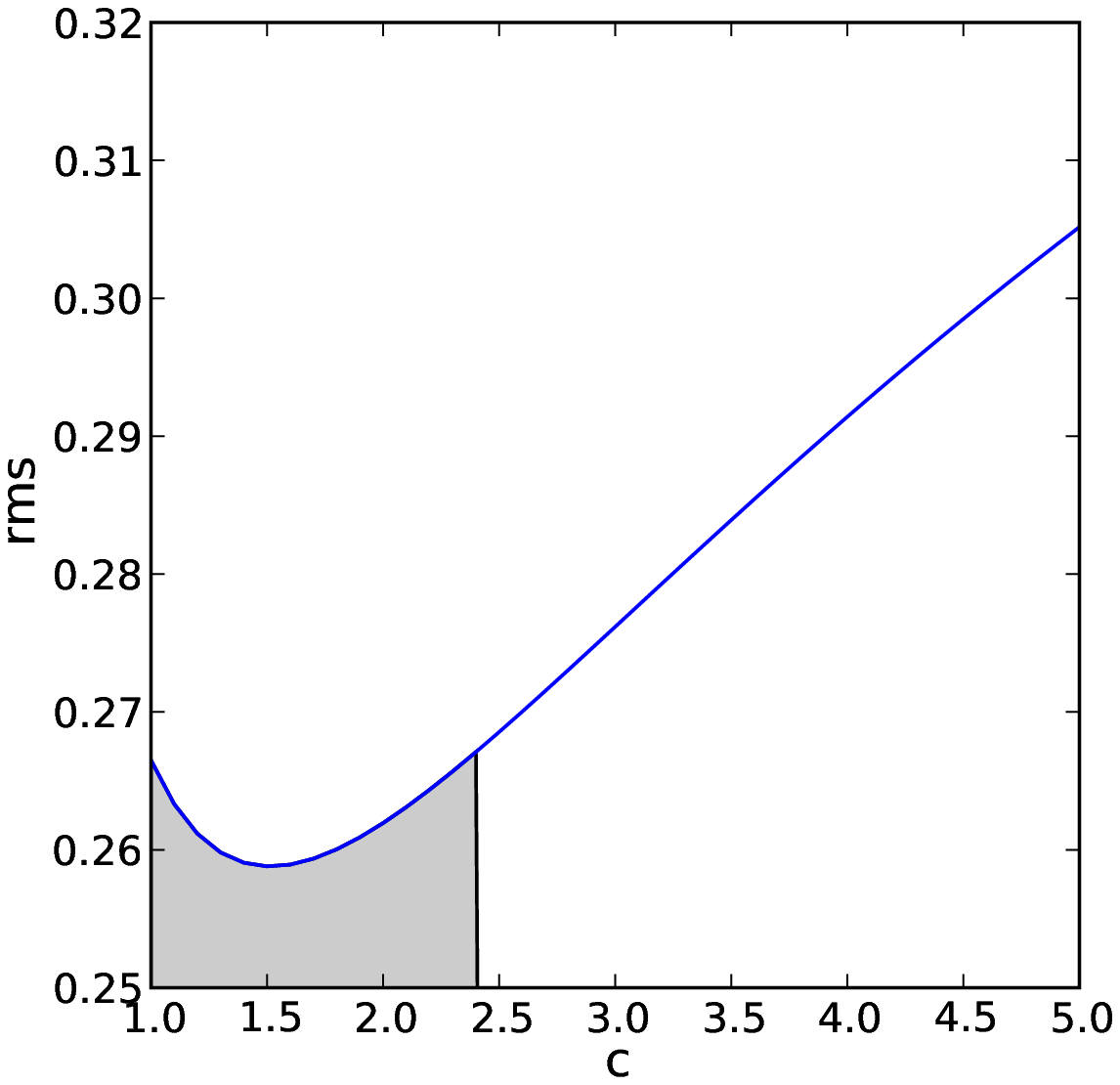}
 \includegraphics{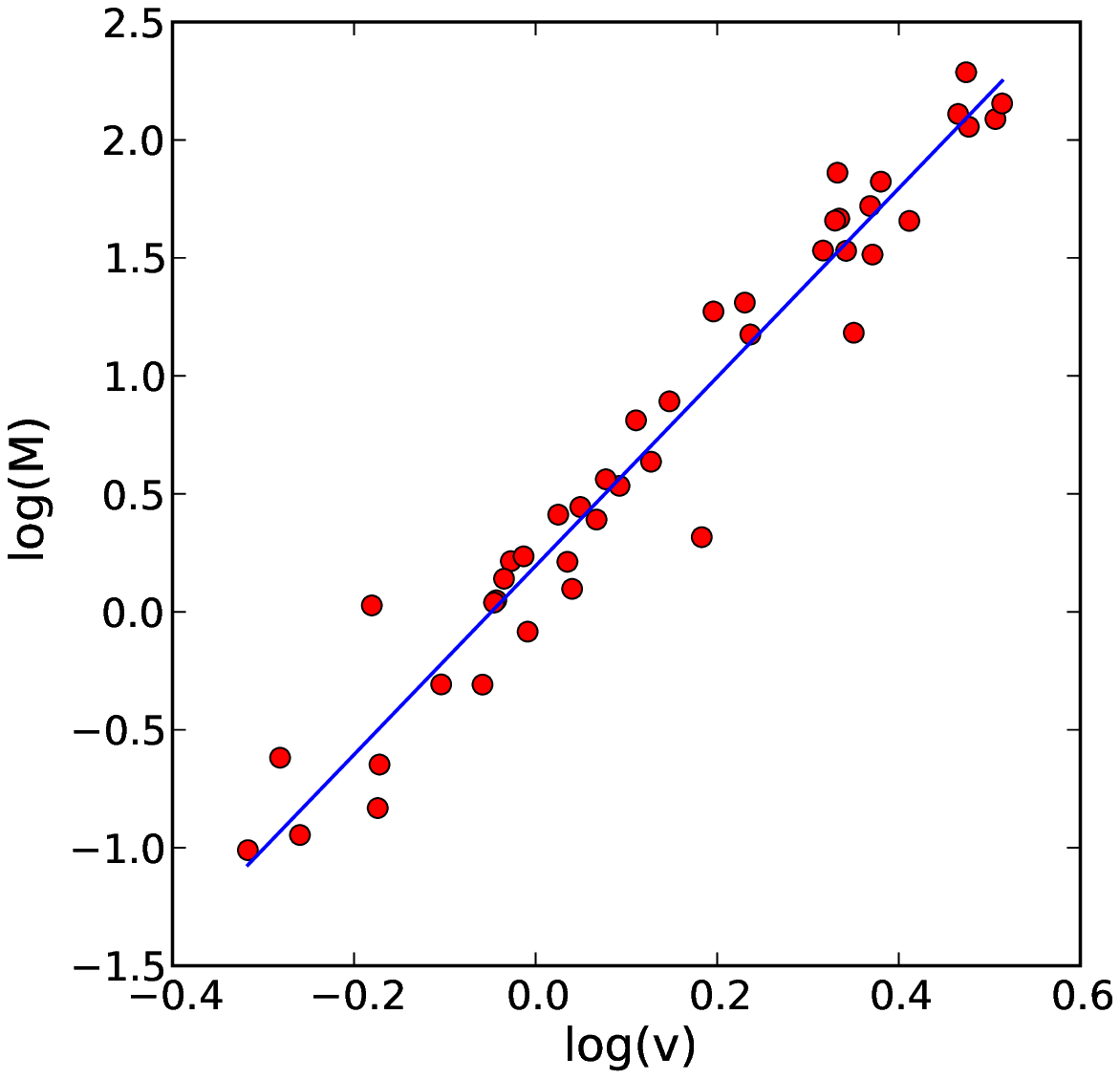}
\includegraphics{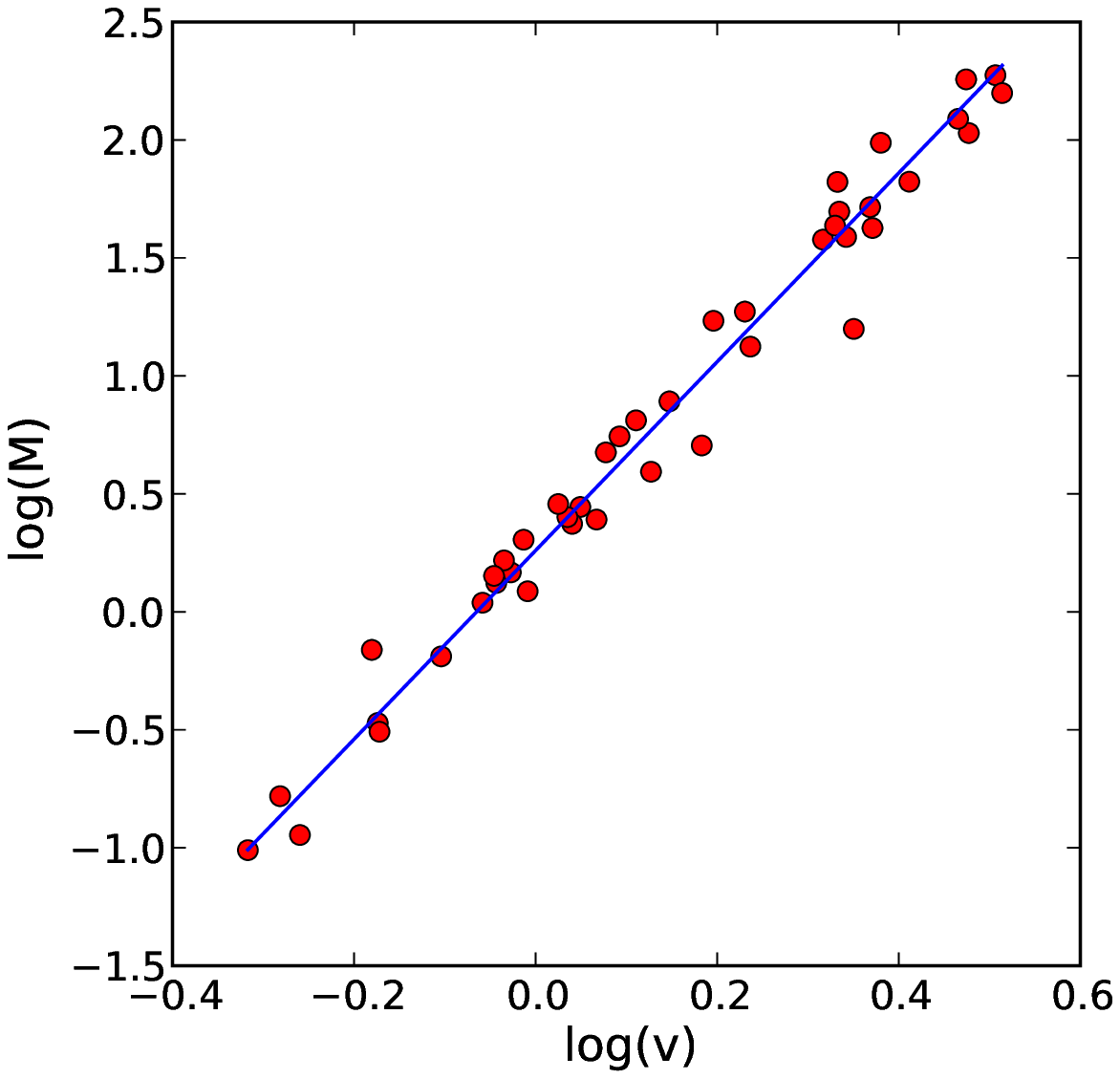}
}	
 % ty.eps: 1048592x1048592 pixel, 300dpi, 8878.08x8878.08 cm, bb=
 \caption{\textit{First sample}.\textit{Left}: Root mean square of the linear least fit square of the Tully-Fisher relation as a function of c. The scatter of the Tully-Fisher relation is a minimum for c=1.5. \textit{Middle}: The baryonic Tully-Fisher relation for c=1.5 whatever the galaxy (the masses and velocity are given in the system unit G=1). \textit{Right}: The baryonic Tully-Fisher relation where the values of c are determined by the rotation curve analysis; they are different from one galaxy to the other. }
\label{fig:tfs1}
\end{figure*}

\begin{figure*}[!]
 \centering
 
\includegraphics[width=6.5cm]{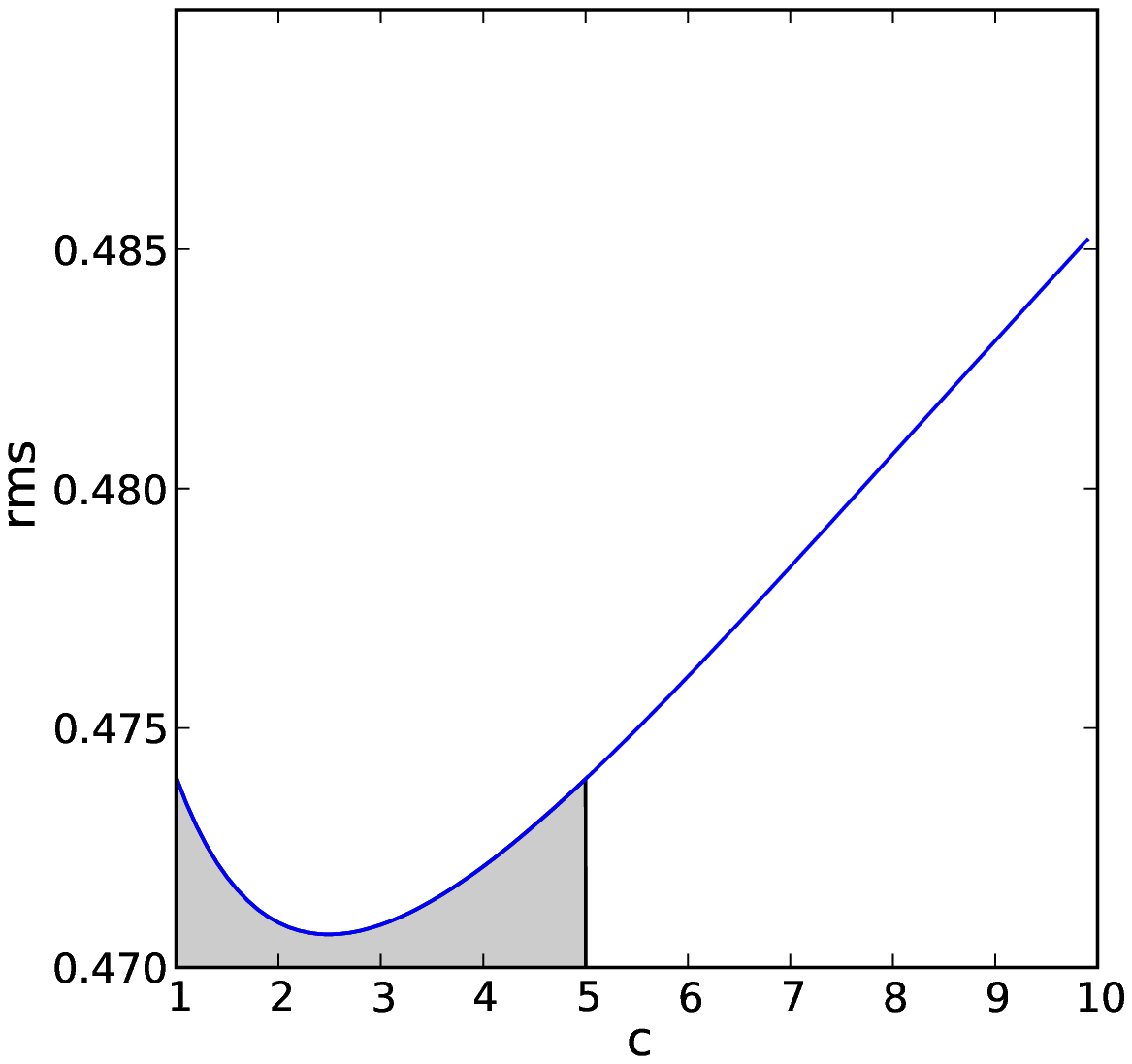}
\includegraphics[width=6.5cm]{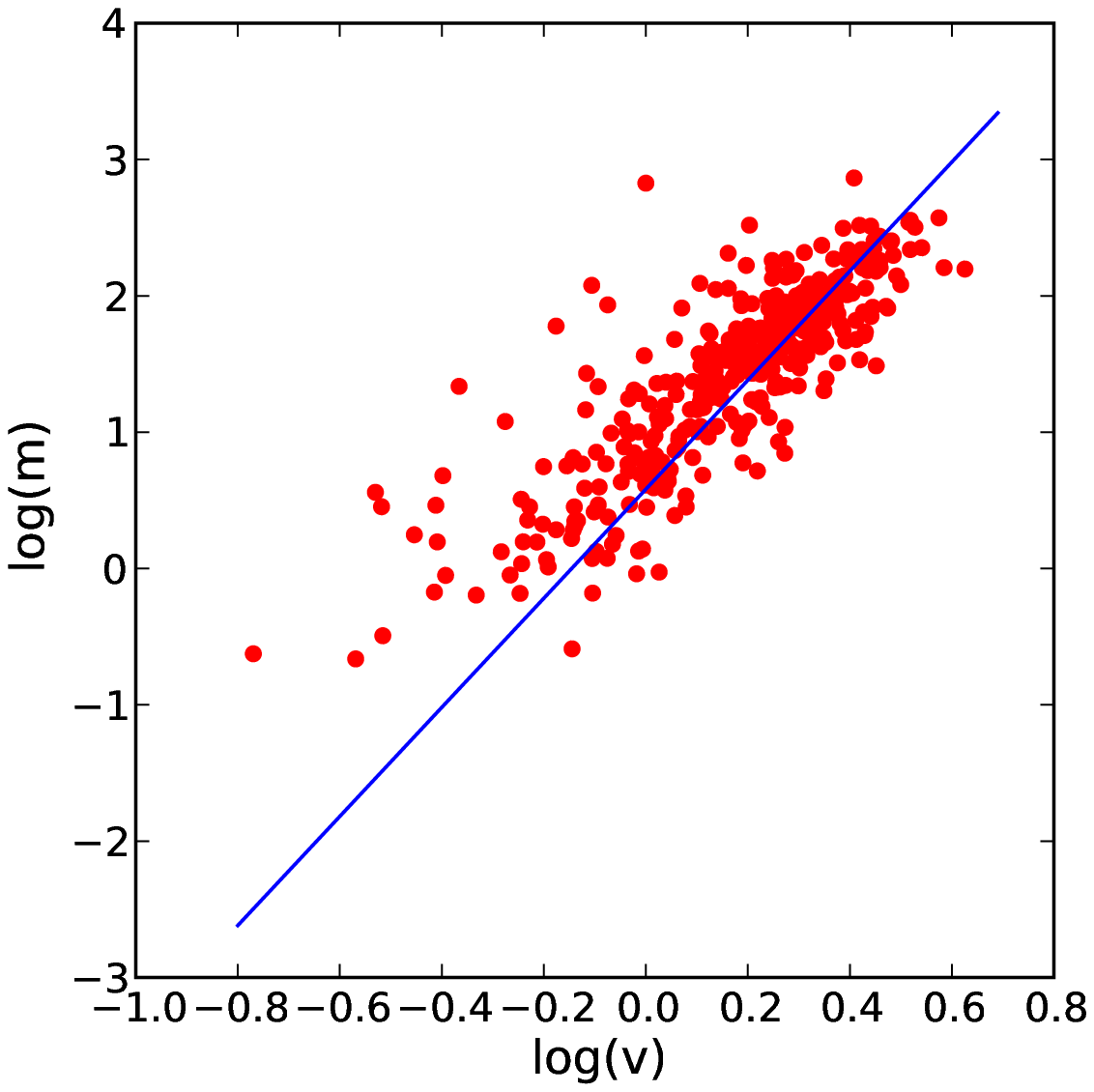}
	
 % ty.eps: 1048592x1048592 pixel, 300dpi, 8878.08x8878.08 cm, bb=
 \caption{\textit{Second sample. Left}: Rms of the linear least square fit of the Tully-Fisher relation as a function of c. The minimum is obtained for c=2.6. \textit{Right}: The baryonic Tully-Fisher relation including a dark-baryon component, c=2.6 whatever the galaxy.}
\label{fig:rms}
\end{figure*}

\subsection{Rotation curve analysis}

The minimization of the $\chi^2$ for the rotation curve fits gives the variation of c with respect to $a_0$ for each galaxy. Fig. \ref{fig:rc_ca0} (left) displays the mean over the galaxy sample of the factor c versus the critical acceleration $a_0$. When $a_0$ tends to zero, which is the Newtonian limit in the MOND formulation, the mean proportion of total gas over the atomic one is $c=12\pm 4$. In this limit all the dark matter needed to fit the rotation curve is in the form of a dark gas disc. This result is in agreement with the one found by Hoekstra et al. (2001). They argue for $c=7$ which is the most common value in their frequency histogram but their mean value is larger ($c=11$, see the error bar in Fig. \ref{fig:rc_ca0}, left). On the other hand for $a_0=1.2 \times 10^{-10}m.s^{-2}$, the actual value of the critical acceleration of MOND, the mean c is equal to 1, which means that there is no need for additional baryons to fit the rotation curve. Thus the critical acceleration $a_0$ increases as the fraction of dark baryons decreases. This last plot gives the general order of magnitude of the values $a_0$ and c.
Now to determine if there is a preferred (universal) value of $a_0$, if some dark baryons are taken into account, we plot a histogram of $a_0$ given by the fit. It is represented in the middle of Fig. \ref{fig:rc_ca0}. It can be seen that the critical acceleration common to the sample peaks at $a_0=0.96\pm 0.39\times 10^{-10}$ m.s$^{-2}$. Then, if we look at the histogram of c in Fig \ref{fig:rc_ca0}, right,  it shows that the majority of the galaxy in the sample ($75\%$) could contain some dark baryons in the proportion $1.5<c<6$, while a few percent of galaxies have a scale-factor of 1, equivalent to a model without dark baryons. The mean value of the scale-factor is $c=3.2\pm2.4$.
Fig. \ref{fig:datacr} displays the different components of the rotation curves fitted by MOND with a fraction of dark baryons.

\subsection{Tully-Fisher relation}

\paragraph{First sample.}
Fig. \ref{fig:tfs1} (left) shows the evolution of the root mean square of the Tully-Fisher relation fit with respect to the fraction between total and atomic gas, c. 
The factor c which optimizes the scatter of the relation is found near $c=1.5$. This value is lower than for the rotation curve method. But this method takes into account only the global properties of the galaxies, so it is a statistical study that necessitates a large number of candidates. In that case we have only 43 galaxies. This is why we analyzed a second sample made of several hundred of galaxies. However note that if we look at the maximum value of c for which the root mean square (rms) of the fit is still better than when there are no dark baryons, we find $c<2.5$ (see the gray region in Fig. \ref{fig:tfs1}, left), which is more in accordance with the results of the fisrt method.

The baryonic Tully-Fisher relation where c is fixed to 1.5, whatever the galaxy, is plotted in Fig \ref{fig:tfs1} (middle). In this case, the root mean square is 0.25. To compare this with the first method, the same baryonic Tully-Fisher relation is plotted in Fig \ref{fig:tfs1} (right) but the values of c are the ones derived from the individual rotation curve analysis, thus different in each galaxy. This is why the scatter is lower than in the previous plot; the root mean square is $0.14$.

\paragraph{Second sample.}
With a larger sample of galaxies (n=576) selected from the GoldMine database, we performed the same study on the baryonic Tully-Fisher relation. This time, the factor that optimizes the scatter of the relation is  $c=2.6\pm2.5$. The uncertainty corresponds to the range of c where the rms is better than if no additional dark baryons are considered (Fig. \ref{fig:rms}). Moreover, the vertical offset of the relation gives  the new value of the critical acceleration of MOND if some dark baryons are considered : $a_0=0.85 \pm 0.35 \times 10^{-10}$ m.s$^{-2}$. This analysis of the Tully-Fisher relation is well  in agreement with the results derived from the rotation curve fits.

\begin{table}
\centering
\begin{tabular}{lcc|lcc}
\hline
\hline
Name & $(M_\star/L)_B$ & c & Name & $(M_\star/L)_B$ & c \\
\hline

ddo154 &   0.1  &  1.5   &   n4258  &    4.1  &   4.5  \\
ddo168 &   0.1  &  1.5   &   n4395  &    3.0  &   2.5  \\
ddo170 &   1.2  &  1.0   &   n45       &    3.5  &   2.0 \\
i2574     &   0.1  &  2.5   &   n4725  &    7.7  &   3.0  \\
m33       &   2.6  &  3.0   &   n5033  &    3.3  &   2.0  \\
n1003   &   0.5  &  1.5   &   n5055  &    3.8  &   3.0  \\
n1560   &   0.2  &  2.0   &   n5371  &    5.6  &   5.0  \\
n2366   &   0.1  &  1.0   &   n55       &    0.1  &   3.5  \\
n2403   &   2.1  &  1.5   &   n5533  &    6.1  &   1.0  \\
n24       &   2.1  &  8.0    &   n5585  &    0.8  &   2.0  \\
n247     &   1.2  &  4.0   &   n5907   &    5.5  &  10.  \\
n253     &   3.0  &  3.5   &   n6503   &    3.0  &   6.0  \\
n2841  &   5.1  &  8.5   &   n6674   &    7.5   &   2.5  \\
n2903  &   2.7  &  3.0   &   n6946   &    2.0   &   3.0  \\
n2915  &   5.8  &  1.0   &   n7331   &    3.3  &   8.0  \\
n2998  &   3.2  &  1.0   &   n7793   &    1.5  &   1.5  \\
n300    &   0.6  &  2.0   &   n801      &    2.4 &   5.0  \\
n3109  &   0.1  &  2.0   &   n925     &    1.1  &   2.5  \\
n3198  &   2.8  &  2.5   &   u128     &    6.0  &   1.5  \\
n3726  &   1.7  &  2.5   &   u2259   &    8.6  &   3.5  \\
n4203  &   0.9  & 10.    &   u2885   &    6.4  &   3.0  \\
n4242  &   0.9  &  3.5   &                  &           &        \\
\hline
\end{tabular}

\caption{c is the scale-factor between the total gas (atomic and molecular) and the atomic gas. $(M_\star/L)_B$ is the mass to light ratio in the B-band.}
\label{tab:param}

\end{table}

\section{Discussion \& conclusion }

We revisit the mass discrepancy-acceleration relation, at the origin of the motivation for MOND. 
In the frame of this modified gravity theory, this relation is a direct observation, which quantifies
the critical acceleration and determines the interpolation function $\mu$. 
In the frame of Newtonian gravity and dark matter theory, we have to understand the meaning of this 
strong correlation, which is related to both the Tully-Fisher and virial relation.
Until now, only fine tuning, involving baryonic physics and supernovae feedback, has been invoked as
an interpretation.

We then consider for the first time how the presence of dark baryons in galaxies
could be made compatible with the MOND phenomenology.
Only a small fraction of all missing baryons can be present in galaxies,
to avoid overpredicting the observed rotation curves. If dark baryons are present in 
cosmic filaments in the form of cold gas, it is unavoidable to find a small fraction of them
in galaxies.
We show how the presence of these dark baryons in galaxy discs can reduce the critical
acceleration $a_0$ of MOND. We quantify the best pair ($a_0$, $c$) with $c$ being the scale-factor between the total gas and the atomic gas in a galaxy. 
The main result is derived from the rotation curve analysis which involves galaxies well resolved with their luminosity profiles and HI profiles.  More galaxies with high-quality data should be included in the sample, especially the late-type galaxies which are very constraining. 
We applied another method based on the minimization of the scatter of the Tully-Fisher relation. We use two different samples, one is the same as for the rotation curve fits (with 43 galaxies); the other contains a larger number of galaxies (576 candidates). Both methods with both samples conclude that the presence of dark baryons in the galaxy disc are compatible with MOND. We find the scale factor between the total gas and atomic gas to be  $c=3_{-2}^{+3}$ (by compiling all the results). This factor is not expected to be universal because of invidual  galaxy histories. On the other hand, we show that the critical acceleration of MOND which differentiates the Newtonian regime from the mondian regime should be slightly reduced from the actual value usually used. In this work the critical acceleration of MOND, if some dark baryons are taken into account, is estimated to be$a_0=0.96\pm 0.39\times 10^{-10}$ m.s$^{-2}$.

Our results on the Tully-Fisher relation are in agreement with those of Pfenniger \& Revaz (2005) and Begum et al. (2008), where the error bars  are large because of the difficulty
in determining with very high precision the visible baryonic mass. They find respectively  $c=3_{-2}^{+9}$ and $c=9_{-7}^{+19}$.
But we interpret the Tully-Fisher relation conforming to MOND, and we fix its slope to 4. 
It assumes that the velocity used in the Tully-Fisher relation corresponds to the constant circular 
velocity in the deep MOND regime. It is possible that the velocity has not yet reached the plateau 
of the MOND regime. This effect will tend to decrease the value of the slope because these samples of 
galaxies contain essentially late-types. Their rotation curves are still increasing  in the outer measured radius, and the velocity observed will always be lower than or equal to the 
asymptotic velocity.

Let us add that a factor $c$ larger than 7 appears less likely from a dynamical point of view 
(Revaz et al. 2008). With such a large factor, the total mass in gas (HI and dark) is about the 
same order as that in the stellar disc, at a redshift z=0. In this case, galaxy discs 
should be too cold and unstable. Note also that the proportion of dark baryons estimated 
in this work corresponds to a quantity within ten times the typical scale length of the galaxies, due to the extent of the rotation curves.

\begin{figure*}
 \centering
  \resizebox{\hsize}{!}{\includegraphics{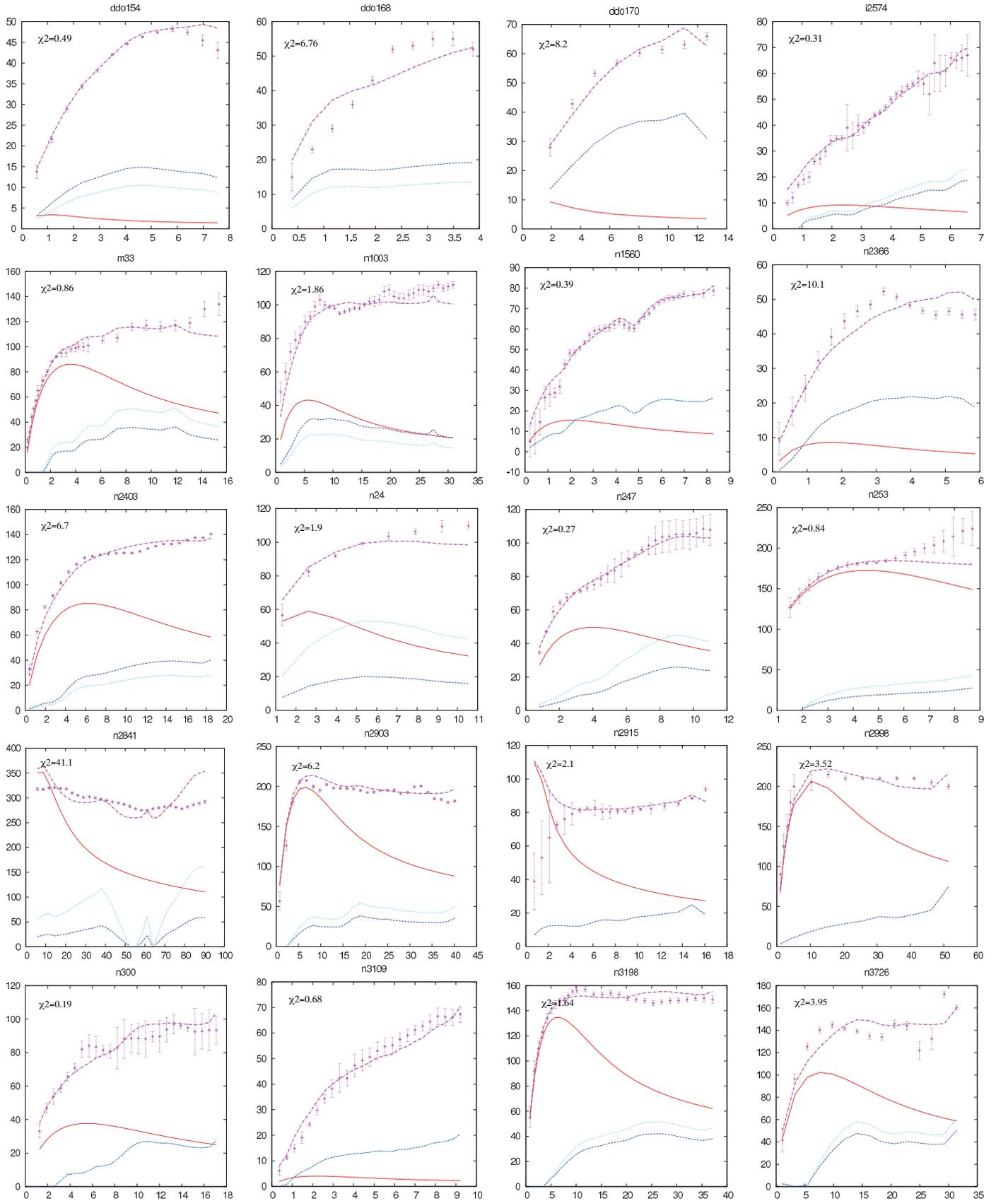}}
\caption{MOND rotation curve fits: the solid line (red) represents the stellar disc, the atomic gas corresponds 
to the dash line (blue), the dot line (cyan) is the cold and dark molecular gas. The total modelled rotation curve is 
in dot-dashed (magenta) while the observed HI velocity corresponds to the symbols and error-bars.}
   \label{fig:datacr}
\end{figure*}

\begin{figure*}
 \centering
  \resizebox{\hsize}{!}{\includegraphics{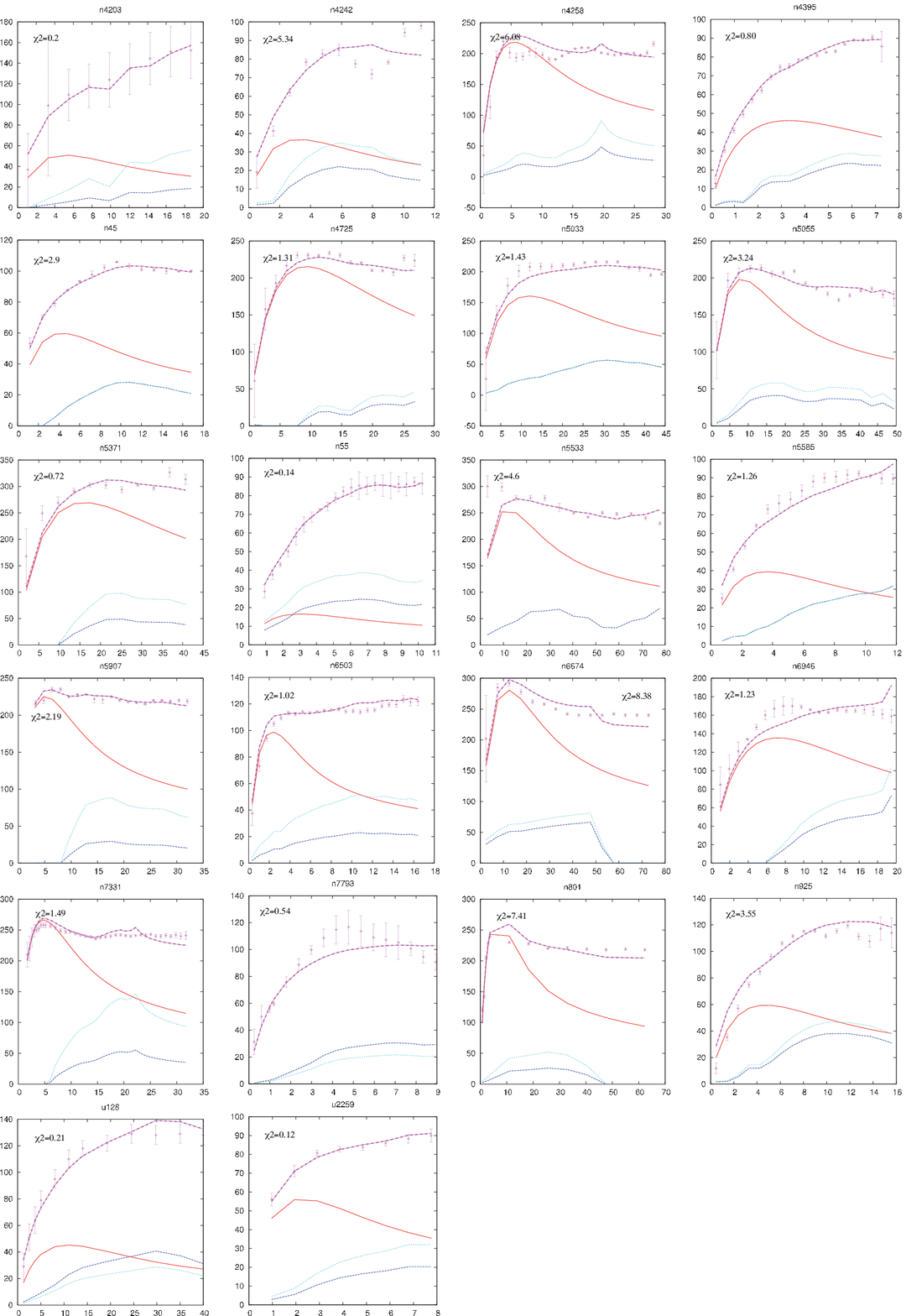}}
\caption{Continued.}
\end{figure*}

%%%%%%%%%%%%%%%%%%%%%%%% acknowledgments
\begin{acknowledgements}
We acknowledge the use of the HyperLeda database (http://leda.univ-lyon1.fr),
and of the GOLD Mine Database (Gavazzi et al. 2003).
This research has made use of the NASA/IPAC Extragalactic Database (NED) which is operated by the Jet Propulsion 
Laboratory, California Institute of Technology, under contract with the National Aeronautics and Space Administration.
We thank the referee for very helpful remarks that improved the paper.
\end{acknowledgements}
%%%%%%%%%%%%%%%%%%%%%%%%%%%%%%%%%%%%%

%%%%%%%%%%%%%% references %%%%%%%%%%%%%%%%
{} 
%%%%%%%%%%%%%%%%%%%%%%%%%%%%%%

\end{document}